\documentclass[12pt]{article}
\usepackage{epsfig}
\newcommand{\be}{\begin{equation}}
\newcommand{\ee}{\end{equation}}
\newcommand{\bqa}{\begin{eqnarray}}
\newcommand{\eqa}{\end{eqnarray}}
\newcommand{\pslash}{\slash\hspace{-0.55em}}
\newcommand{\as}{\alpha_{\mathrm{s}}}

\begin{document}

\begin{center}
{\LARGE $B \rightarrow \eta_c K(\eta_c^\prime K)$ decays in QCD factorization}\\[0.8cm]
{\large Zhongzhi Song$~^{(a)}$, Ce Meng$~^{(a)}$
and~Kuang-Ta Chao$~^{(b,a)}$}\\[0.5cm]
{\footnotesize (a)~Department of Physics, Peking University,
 Beijing 100871, People's Republic of China}

{\footnotesize (b)~China Center of Advanced Science and Technology
(World Laboratory), Beijing 100080, People's Republic of China}
\end{center}

\vspace{0.5cm}
\begin{abstract}
We study the exclusive decays of $B$ meson into pseudoscalar
charmonium states $\eta_c$ and $\eta_c^\prime$ within the QCD
factorization approach and find that the nonfactorizable
corrections to naive factorization are infrared safe at
leading-twist order. The spectator interactions arising from the
kaon twist-3 effects are formally power-suppressed but chirally
and logarithmically enhanced. The theoretical decay rates are too
small to accommodate the experimental data. On the other hand, we
compare the theoretical calculations for $J/\psi, \psi^\prime$,
and $\eta_c, \eta_c^\prime$, and find that the predicted relative
decay rates of these four states are approximately compatible with
experimental data.

\vspace{0.5cm}
\end{abstract}

\section{Introduction}
Exclusive decays of $B$ meson to charmonium are important since
those decays e.g. $B \to J/\psi K$ are regarded as the golden
channels for the study of CP violation in $B$ decays. However,
quantitative understanding of these decays is difficult due to the
strong-interaction effects. It is conjectured physically that
because the size of the charmonium is small$(\sim 1/{\as
{m_\psi}})$ and its overlap with the $(B, K)$ system is
negligible\cite{BBNS1}, the same QCD-improved factorization method
as for $B\to\pi\pi$\cite{BBNS2,BBNS3} can be used for $B \to
J/\psi K$ decay. Indeed, for this channel the explicit
calculations \cite{chay,cheng} show that the nonfactorizable
vertex contribution is infrared safe and the spectator
contribution is perturbatively calculable at twist-2 order. This
small size argument for the applicability of QCD factorization for
the charmonia is intuitive, but it needs verifying for charmonium
states other than the $J/\psi$ and $\psi^\prime(\psi(2S))$.

In our previous paper\cite{song}, we studied the $B \rightarrow
\chi_{cJ} K (J=0,1)$ decays within the QCD factorization approach,
and found that for $B \rightarrow \chi_{c1} K$ decay, the
factorization breaks down due to logarithmic divergences arising
from nonfactorizable spectator interactions even at twist-2 order,
and that for $B\rightarrow \chi_{c0} K$ decay, there are infrared
divergences arising from nonfactorizable vertex corrections as
well as logarithmic divergences due to spectator interactions even
at leading- twist order.

Experimentally, for the pseudoscalar charmonium state $\eta_c$,
the $B \rightarrow \eta_c K$ decay has been observed by
CLEO\cite{cleo}, BaBar\cite{babar}, and Belle\cite{belle1} with
relatively large branching fractions. Moreover, very recently the
$\eta_c^\prime(\eta_c(2S))$ meson has also been observed in the $B
\rightarrow \eta_c^\prime K$ decay by Belle\cite{belle2}. So, it
is interesting to compare the predictions of these decay modes
into pseudoscalar charmonium based on the QCD factorization
approach with the experimental data to further test the
applicability of QCD factorization to $B$ meson exclusive decays
to charmonium states.

\section{$B\rightarrow \eta_c K$ decay within QCD
factorization}\label{s2}
 We now consider $\overline{B}\rightarrow
\eta_c K$ decay. The effective Hamiltonian for this decay mode is
written as\cite{BBL}
 \be
H_{\mathrm{eff}} = \frac{G_F}{\sqrt{2}} \Bigl( V_{cb} V_{cs}^*
(C_1 {\cal O}_1 +C_2 {\cal O}_2 ) -V_{tb} V_{ts}^* \sum_{i=3}^{10}
C_i {\cal O}_i \Bigr),
 \ee
where $C_i$ are the Wilson coefficients and the relevant operators
${\cal O}_i$ in $H_{\mathrm{eff}}$ are given by
 \bqa
&& {\cal O}_1=(\overline{s}_{\alpha} b_{\beta})_{V-A} \cdot
(\overline{c}_{\beta} c_{\alpha})_{V-A},\qquad\qquad~ {\cal
O}_2=(\overline{s}_{\alpha} b_{\alpha})_{V-A} \cdot
(\overline{c}_{\beta} c_{\beta})_{V-A},
 \nonumber\\
&& {\cal O}_{3(5)}=(\overline{s}_{\alpha} b_{\alpha})_{V-A} \cdot
\sum_q (\overline{q}_{\beta} q_{\beta})_{V-A(V+A)},~ {\cal
O}_{4(6)}=(\overline{s}_{\alpha} b_{\beta})_{V-A} \cdot \sum_q
(\overline{q}_{\beta} q_{\alpha})_{V-A(V+A)},
\\
&& {\cal O}_{7(9)}={3\over 2}(\overline{s}_{\alpha}
b_{\alpha})_{V-A} \cdot \sum_q e_q (\overline{q}_{\beta}
q_{\beta})_{V+A(V-A)},~ {\cal O}_{8(10)}={3\over
2}(\overline{s}_{\alpha} b_{\beta})_{V-A} \cdot \sum_q e_q
(\overline{q}_{\beta} q_{\alpha})_{V+A(V-A)}.\nonumber
 \eqa

To calculate the decay amplitude, we introduce the $\eta_c$ decay
constant as
 \be
 \langle \eta_c(p) | \overline{c}
(0) \gamma_{\mu}\gamma_5 c(0) |0\rangle = -if_{\eta_c} p_{\mu},
\label{cons}
 \ee
where $f_{\eta_c}$ is the $\eta_c$ decay constant which can be
estimated from the QCD sum rules or potential models. The
leading-twist light-cone distribution amplitude of $\eta_c$ is
then expressed compactly as
 \bqa
  \langle \eta_c
(p)| \overline{c}_{\alpha} (z_2) c_{\beta} (z_1) |0\rangle &=&
\frac{i f_{\eta_c}}{4} \int_0^1 du \cdot e^{i(u p\cdot z_2 + (1-u)
p \cdot z_1)} \bigl[( \pslash{p}+m_{\eta_c}) \gamma_5
\bigr]_{\beta \alpha} \phi_{\eta_c}(u),
  \label{lcda}
 \eqa
where $u$ and $1-u$ are respectively the momentum fractions of the
$c$ and $\bar c$ quarks inside the $\eta_c$ meson, and the wave
function $\phi_{\eta_c}(u)$ for $\eta_c$ meson is symmetric under
$u\leftrightarrow 1-u$.

As for the kaon light-cone distribution amplitudes, we will follow
Ref.~\cite{BBNS3}to choose
  \bqa
    &&\langle K(p)|\bar s_\beta(z_2)\,d_\alpha(z_1)|0\rangle
    \nonumber\\
   &&= \frac{i f_K}{4} \int_0^1\! dx\,
    e^{i(x\,p\cdot z_2+(1-x)\,p\cdot z_1)}
    \left\{ \pslash{p}\,\gamma_5\,\phi_K(x)
    - \mu_K\gamma_5 \left( \phi_K^p(x) - \sigma_{\mu\nu}\,p^\mu (z_2-z_1)^\nu\,
    \frac{\phi_K^\sigma(x)}{6} \right) \right\}_{\alpha\beta} . \quad
\label{kaon}
  \eqa
where $x$ and $1-x$ are respectively the momentum fractions of the
$s$ and $\bar d$ quarks inside the $K$ meson. The asymptotic limit
of the leading-twist distribution amplitude is
$\phi_K(x)=6x(1-x)$. We also use the asymptotic forms
$\phi_K^p(x)=1$ and $\phi_K^\sigma(x)=6x(1-x)$ for the kaon
twist-3 two-particle distribution amplitudes. The
chirally-enhanced factor is written as
$r_\chi^K=2\mu_K/m_b=2m_K^2/m_b(m_s+m_d)$ which is formally of
order $\Lambda_{\mathrm{QCD}}/m_b$ but numerically close to unity.

In the naive factorization, we neglect the strong interaction
corrections and the power corrections in
$\Lambda_{\mathrm{QCD}}/m_b$. Then the decay amplitude is written
as \bqa
 iM_0=i f_{\eta_c} m_B^2 F_0 (m_{\eta_c}^2) \frac{G_F}{\sqrt{2}}
 \Bigl[ V_{cb} V_{cs}^* (C_2
+\frac{C_1}{N_c} ) -V_{tb} V_{ts}^* (C_3 + \frac{C_4}{N_c} -C_5
-\frac{C_6}{N_c}) \Bigr], \label{tree}
 \eqa
where $N_c$ is the number of colors. We do not include the effects
of the electroweak penguin operators since they are numerically
small. The form factors for $\overline{B} \rightarrow K$ are given
as
  \bqa
\langle K(p_K) | \overline{s} \gamma_{\mu} b| B(p_B)\rangle=\Bigl[
(p_B +p_K)_{\mu} -\frac{m_B^2-m_K^2}{p^2} p_{\mu} \Bigr] F_1 (p^2)
+ \frac{m_B^2-m_K^2}{p^2} p_{\mu} F_0 (p^2),
 \label{vmu2}
  \eqa
where $p= p_B -p_K$ is the momentum of $\eta_c$ with $p^2 =
m_{\eta_c}^2$ and we will neglect the kaon mass for simplicity.

\begin{figure}[t]
\begin{center}
\vspace{-3.5cm}
\includegraphics[width=14cm,height=18cm]{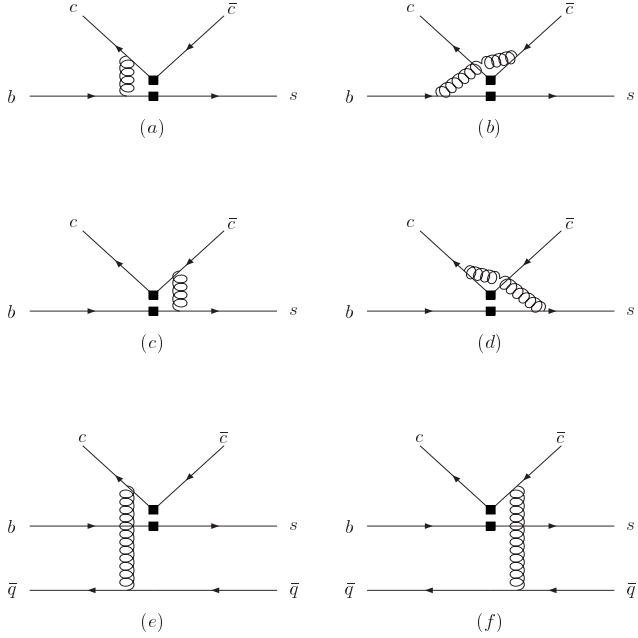}
\vspace{-5.5cm}
\end{center}
\caption{ Feynman diagrams for nonfactorizable corrections to
$\overline{B} \rightarrow \eta_c K$ decay.} \label{fvs}
\end{figure}

As we can see easily in Eq.~(\ref{tree}), this amplitude is
unphysical because the Wilson coefficients depend on the
renormalization scale $\mu$ while the decay constant and the form
factors are independent of $\mu$. This is the well known problem
with the naive factorization. However, if we include the order
$\alpha_s$ corrections, it turns out that the $\mu$ dependence of
the Wilson coefficients is cancelled and the overall amplitude is
insensitive to the renormalization scale. Taking the
nonfactorizable order $\alpha_s$ strong-interaction corrections in
Fig.~\ref{fvs} into account, the full decay amplitude for $
\overline{B} \rightarrow \eta_c K$ within the QCD factorization
approach is written as
 \be
  iM = i f_{\eta_c} m_B^2
F_0 (m_{\eta_c}^2) \frac{G_F}{\sqrt{2}}\Bigl[ V_{cb} V_{cs}^* a_2
-V_{tb} V_{ts}^* (a_3 - a_5) \Bigr], \label{full}
 \ee
where the coefficients $a_i$ ($i=2,3,5$) in the naive dimension
regularization(NDR) scheme are given by  \bqa
 && a_2=C_2 +\frac{C_1}{N_c} +\frac{\alpha_s}{4\pi}
\frac{C_F}{N_c} C_1 \Bigl( -18 +12\ln \frac{m_b}{\mu} + f_I +
f_{II}
\Bigr), \nonumber \\
&&a_3=C_3 +\frac{C_4}{N_c} +\frac{\alpha_s}{4\pi} \frac{C_F}{N_c}
C_4 \Bigl( -18 +12 \ln \frac{m_b}{\mu} +f_I + f_{II}
\Bigr),   \label{ai}\\
&&a_5=C_5 +\frac{C_6}{N_c} -\frac{\alpha_s}{4\pi} \frac{C_F}{N_c}
C_6 \Bigl( -6 +12 \ln \frac{m_b}{\mu} +f_I + f_{II}
\Bigr).\nonumber
  \eqa

The function $f_I$ in Eq.(\ref{ai}) is calculated from the four
vertex corrections (a,b,c,d) in Fig.~\ref{fvs} and it reads
 \bqa
  f_I&=&\int_0^1 du\  \phi_{\eta_c} (u) \Bigl[ \frac{3(1-2u)}{1-u}\ln[u]+3( \ln
  (1-z)-i \pi)-\frac{2z (1-u)}{1-z u}-\frac{2 u z(\ln[1 - z]-i \pi )}{1 - (1 - u) z}
  \nonumber \\&-&\frac{ u^2 z^2 (\ln[1 - z]-i \pi )}{(1 - (1 - u) z)^2}
 +u z^2  \ln[u z]\bigl(\frac{u}{(1-(1 - u) z)^2}-\frac{1-u}{(1-u z)^2}\bigr)\nonumber \\
 &+&2 u z \ln[u z] \bigl(\frac{1}{1-(1 - u) z}
 -\frac{1}{1-u z}\bigr) \Bigr],
 \eqa
where $z=m_{\eta_c}^2/m_B^2$, and we have already symmetrized the
result with respect to $u\leftrightarrow 1-u$.

The function $f_{II}$ in Eq.(\ref{ai}) is calculated from the two
spectator interaction diagrams (e,f) in Fig.~\ref{fvs} and it is
given by
 \bqa
  f_{II}
= \frac{4\pi^2}{N_c} \frac{f_K f_B}{m_B^2 F_0 (m_{\eta_c}^2)}
\int_0^1 d\xi \frac{\phi_B (\xi)}{\xi} \int_0^1 du
\frac{\phi_{\eta_c}(u)}{u} \int_0^1 \frac{dx}{x} [\phi_K
(x)+\frac{2\mu_K \phi_K^p (x) }{m_b (1-z)}], \label{fII}
 \eqa
 where $\phi_B$ is the light-cone wave functions for the $B$ meson.
 The spectator contribution depends on the wave function
$\phi_B$ through the integral \be
 \int_0^1 d\xi \frac{\phi_B (\xi)}{\xi} \equiv
\frac{m_B}{\lambda_B}.
 \ee
Since $\phi_B (\xi)$ is appreciable only for $\xi$ of order
$\Lambda_{\mathrm{QCD}}/m_B$, $\lambda_B$ is of order
$\Lambda_{\mathrm{QCD}}$. We will follow Ref.~\cite {BBNS3} to
choose $\lambda_B\approx 300$ MeV in the numerical calculation.

\begin{table}[t]
\begin{center}
\begin{tabular}{ c | c c c c c c }
   \hline
    &$C_1$ & $C_2$ & $C_3$ & $C_4$
 & $C_5$ & $C_6$ \\
 \hline
  LO & 1.144 & -0.308& 0.014 & -0.030 & 0.009 & -0.038  \\
  NDR & 1.082 & -0.185 & 0.014 & -0.035 & 0.009 & -0.041 \\
 \hline
 \end{tabular}
\caption{ {Leading-order(LO) and Next-to-leading-order(NLO) Wilson
coefficients in NDR scheme(See Ref.\cite{BBL}) with $\mu=4.4$ GeV
and $\Lambda^{(5)}_{\overline{\rm MS}}=225$ MeV.}}
 \label {wilson}
\end{center}
\end{table}

There is an integral in Eq.~(\ref {fII}) arising from kaon twist-3
effects, which will give logarithmic divergence. Following
Ref.~\cite {BBNS3}, we treat the divergent integral as an unknown
parameter and write
  \be
\int_0^1 dx \frac{\phi_K^p (x)}{x}=\int_0^1 \frac{dx}{x}=X_H,
 \ee
where $\phi_K^p (x)=1$ is used for the kaon twist-3 light-cone
distribution amplitude. We will choose
$X_H=\ln(m_B/\Lambda_{QCD})\approx 2.4$ as a rough estimate in our
calculation.

For numerical analysis, we choose $F_0 (m_{\eta_c}^2) = 0.41$\cite
{ball} and use the following input parameters:
 \bqa
&&m_b=4.8 \ \mbox{GeV}, \ \ m_B=5.28 \ \mbox{GeV}, \ \ m_{\eta_c}
=3.0 \ \mbox{GeV}, \nonumber \\ &&f_{\eta_c}=350 \
\mbox{MeV}\cite{constant}, \ \ f_B = 180 \ \mbox{MeV}, \ \ f_K =
160 \ \mbox{MeV}.
 \eqa

\begin{table}[tb]
\begin {center}
\begin{tabular}{ c|ccc}
 \hline
 $\phi_{\eta_c}(u)$ &$a_2$&$a_3$&$a_5$ \\   \hline
$6u(1-u)$&0.1043-0.0684i&0.0045+0.0022i&-0.0035-0.0026i \\

$\delta (u-1/2)$&0.0792-0.0682i&0.0055+0.0022i&-0.0048-0.0026i \\
\hline
\end{tabular}
\caption{The coefficients $a_i$ at $\mu=4.4$ GeV with different
choices of $\phi_{\eta_c}(u)$.}
 \label{table1}
\end {center}
\end{table}

The asymptotic form of the distribution amplitude
$\phi_{\eta_c}(u)$ is given as $\phi_{\eta_c}(u) = 6u (1-u)$. In
the numerical analysis, we also consider the form
$\phi_{\eta_c}(u) =\delta (u-1/2)$, which comes from the naive
expectation of the distribution amplitude. Although there are
uncertainties associated with the form of the wave function, we
will see shortly that the calculated decay rates are not very
sensitive to the choice of the distribution amplitude. The results
of coefficients $a_i$ are listed in Table.~\ref {table1}.

With the help of these coefficients $a_i$, we calculated the decay
branching ratios. For $\phi_{\eta_c} (u) = 6u(1-u)$,
${\mathrm{Br}} (\overline{B} \rightarrow \eta_c K) = 1.9 \times
10^{-4}$. And for $\phi_{\eta_c}(u) = \delta (u-1/2)$,
${\mathrm{Br}} (\overline{B} \rightarrow \eta_c K) = 1.4 \times
10^{-4}$.

The measured branching ratios are
  \bqa
\mbox{CLEO Collaboration \cite{cleo}: } {\mathrm{Br}} (B^0
\rightarrow \eta_c K^0) &=& (1.09^
{+0.55}_{-0.42})\times 10^{-3},\nonumber\\
\mbox{BaBar Collaboration \cite{babar}: }{\mathrm{Br}} (B^0
\rightarrow \eta_c K^0) &=& (1.06 \pm 0.28)
\times 10^{-3},\\
\mbox{Belle Collaboration \cite{belle1}: }{\mathrm{Br}} (B^0
\rightarrow \eta_c K^0) &=& (1.23 \pm 0.23) \times
10^{-3},\nonumber
  \eqa
which are about seven times larger than our theoretical results.

\section{$B\rightarrow \eta_c^{\prime} K$ decay }\label{s3}
The calculation of the branching ratio for
$\overline{B}\rightarrow \eta_c^{\prime} K$ decay is similar to
that for the $\eta_c$ given above. And we can also get a rough
estimate for the decay rates ratio of $\eta_c^\prime$ to $\eta_c$
in the leading order:
 \be
\frac{{\mathrm{Br}} (B^0\rightarrow \eta_c^\prime
K^0)}{{\mathrm{Br}} (B^0 \rightarrow \eta_c K^0)} \approx
\bigl(\frac{f_{\eta_c^\prime}}{f_{\eta_c}}\bigr)^2
\cdot\bigl[\frac{F_1(m^2_{\eta_c^\prime})}{F_1(m^2_{\eta_c})}\bigr]^2
\cdot\bigl[\frac{m^2_B-m^2_{\eta_c^\prime}}{m^2_B-m^2_{\eta_c}}\bigr]^3
\approx 0.9\times
\bigl(\frac{f_{\eta_c^\prime}}{f_{\eta_c}}\bigr)^2 \approx 0.45,
\label{r1}
  \ee
where we have used $f_{\eta_c^\prime}/f_{\eta_c}\approx
f_{\psi^\prime}/f_{J/\psi}$ with $f_{J/\psi}=400$ MeV,
$f_{\psi^\prime}=280$ MeV, which are determeined from the observed
leptonic decay widths \cite{pdg}; and $F_0(p^2)=(1-p^2/m_B^2)
F_1(p^2)$ with $F_1(m^2_{\eta_c^\prime})=0.81,
F_1(m^2_{\eta_c})=0.58$ \cite{chay, ball}. The ratio in
Eq.(\ref{r1}) will roughly hold even when we include the
$\mathcal{O}(\as)$ corrections, because the $\mathcal{O}(\as)$
corrections are small and the mass difference as well as the wave
function difference between $\eta_c$ and $\eta_c^\prime$ will not
change the values of $a_i$ in Eq.(\ref{ai}) greatly.

The Belle Collaboration has reported the observation of the
$\eta_c^\prime$ in exclusive $B\rightarrow KK_SK^-\pi^+$
decays\cite{belle2}:
  \be
   \frac{{\mathrm{Br}} (B^0\rightarrow \eta_c^\prime K^0)
   {\mathrm{Br}} (\eta_c^\prime \rightarrow K_S K^- \pi^+)}
   {{\mathrm{Br}} (B^0 \rightarrow \eta_c K^0)
   {\mathrm{Br}} (\eta_c \rightarrow K_S K^- \pi^+)}=0.38\pm0.12\pm0.05.
    \label{r2} \ee

As was noted in Ref.\cite{chao}, the hadronic decay branching
fractions for $\eta_c$ and $\eta_c^\prime$ are expected to be
roughly equal for the helicity non-suppressed decay
channels\footnote{It will also be interesting to detect the
helicity suppressed decay channels of $\eta_c$ and $\eta_c^\prime$
in $B$ decays, and to see the differences between the helicity
suppressed (e.g. $\rho\rho, K^* \bar{K}^{*}, \phi\phi$ $p\bar{p}$)
and non-suppressed (e.g. $K\bar K\pi, \eta\pi\pi,
\eta^\prime\pi\pi$) decays of $\eta_c$ and $\eta_c^\prime$. This
will be useful to clarify the helicity suppression mechanism for
the charmonium hadronic decays and the so-called $\rho\pi$ puzzle
in $J/\psi$ and $\psi^\prime$ decays observed by BES and MARKII in
$e^+e^-$ annihilation experiments. For details, see
Refs.\cite{chao, gu}.}. So we have ${\mathrm{Br}} (\eta_c^\prime
\rightarrow K_S K^- \pi^+)\approx {\mathrm{Br}} (\eta_c
\rightarrow K_S K^- \pi^+)$, and then from Eq.(\ref{r2}) we get
 \be
  \frac{{\mathrm{Br}} (B^0\rightarrow
\eta_c^\prime K^0)}{{\mathrm{Br}} (B^0 \rightarrow \eta_c K^0)}
\approx 0.4,
 \label{r3}\ee
which is consistent with the ratio in Eq.(\ref{r1}). However, as
was mentioned above,  because the theoretical decay rate is about
seven times smaller than the the experimental data for
${\mathrm{Br}} (B^0 \rightarrow \eta_c K^0)$, the theoretical
branching fraction will also be about seven times smaller than the
the experimental data for ${\mathrm{Br}} (B^0 \rightarrow
\eta_c^\prime K^0)$.

\section{Discussion}
We have shown that for $B$ decays to $\eta_c$ and $\eta_c^\prime$
the theoretical branching fractions are all about seven times
smaller than the experimental data. However, from Eq.(\ref{r1})
and Eq.(\ref{r3}), we see that the theoretical ratio of the decay
rates of the two states is consistent with experimental data:
 \be
\frac{{\mathrm{Br}} (B^0\rightarrow \eta_c^\prime
K^0)}{{\mathrm{Br}} (B^0 \rightarrow \eta_c K^0)}_{Th.}\approx~
\frac{{\mathrm{Br}} (B^0\rightarrow \eta_c^\prime
K^0)}{{\mathrm{Br}} (B^0 \rightarrow \eta_c K^0)} _{Ex.}.
   \label{r4}\ee

It is also interesting to find that although the theoretical
branching fractions of $B$ meson exclusive decays to $J/\psi$ and
$\psi^\prime$ are both much smaller than the experimental data,
the theoretical ratio of the decay rates of these two states is
also roughly consistent with experimental data\cite{pdg}:
 \be
\frac{{\mathrm{Br}} (B^0\rightarrow \psi^\prime
K^0)}{{\mathrm{Br}} (B^0 \rightarrow J/\psi K^0)} _{Th.}\approx
\bigl(\frac{f_{\psi^\prime}}{f_{J/\psi}}\bigr)^2
\cdot\bigl[\frac{F_1(m^2_{\psi^\prime})}{F_1(m^2_{J/\psi})}\bigr]^2
\cdot\bigl[\frac{m^2_B-m^2_{\psi^\prime}}{m^2_B-m^2_{J/\psi}}\bigr]^3
\approx 0.9\times
\bigl(\frac{f_{\psi^\prime}}{f_{J/\psi}}\bigr)^2\approx 0.45,
\label{r5}\ee \be
 \frac{{\mathrm{Br}} (B^0\rightarrow \psi^\prime
K^0)}{{\mathrm{Br}} (B^0 \rightarrow J/\psi K^0)}_{Ex.}\approx
0.6,
 \label{r6}\ee
where $F_1(m^2_{\psi^\prime})=0.83$ and $F_1(m^2_{J/\psi})=0.61$
are used.

Another interesting observation is that the theoretical ratio of
the branching fractions of $B$ meson exclusive decays to $\eta_c$
and $J/\psi$ is also roughly consistent with experimental
data\cite{cleo,babar, belle1,pdg}:
 \be
  \frac{{\mathrm{Br}}(B^0\rightarrow \eta_c K^0)}{{\mathrm{Br}} (B^0 \rightarrow
J/\psi K^0)} _{Th.}\approx
\bigl(\frac{f_{\eta_c}}{f_{J/\psi}}\bigr)^2
\cdot\bigl[\frac{F_1(m^2_{\eta_c})}{F_1(m^2_{J/\psi})}\bigr]^2
\cdot\bigl[\frac{m^2_B-m^2_{\eta_c}}{m^2_B-m^2_{J/\psi}}\bigr]^3
\approx 1.0 \times \bigl(\frac{f_{\eta_c}}{f_{J/\psi}}\bigr)^2
\approx 0.75,\label{r7}
 \ee
   \be
 \frac{{\mathrm{Br}} (B^0\rightarrow \eta_c K^0)}
 {{\mathrm{Br}} (B^0 \rightarrow J/\psi K^0)}_{Ex.}\approx 1.0 .
 \label{r8}\ee
Eq.(\ref{r7}) also approximately holds when $\mathcal{O}(\as)$
corrections are included.

So, the predicted relative rates of all S-wave charmonium states
$J/\psi, \psi^\prime, \eta_c, \eta_c^\prime$ in the QCD
factorization approach are roughly compatible with data. This has
been shown explicitly above in the leading order approximation,
and even holds when including $\mathcal{O}(\as)$ corrections with
which the calculated decay rates for these four charmonium states
are almost equally smaller than data by a factor of 7-10 though
there are some theoretical uncertainties associated with form
factors\footnote{In our calculation, we have used the relation
$F_0(p^2)=(1-p^2/m_B^2) F_1(p^2)$ derived in Ref.\cite{chay},
which is consistent with the form factors obtained in
Ref.\cite{ball}. This will reduce the effects of uncertainties
arising from form factors on the decay rate ratios. For example,
in Eq.(\ref{r7}) we have used \bqa
(\frac{F_0(m^2_{\eta_c})}{F_1(m^2_{J/\psi})})^2
=(1-m^2_{\eta_c}/m_B^2)^2 (\frac
{F_1(m^2_{\eta_c})}{F_1(m^2_{J/\psi})})^2 \approx
(1-m^2_{\eta_c}/m_B^2)^2=0.46.\nonumber \eqa  This value is close
to that given in Ref.\cite{keum2}, which is the modified version
of Ref.\cite{keum1}. In Ref.\cite{keum2}, the authors discussed
the $B$ decay rate ratio of $\eta_c$ to $J/\psi$ at the leading
order and assumed that $F_0(p^2)$ is a constant and $F_1(p^2)$ has
a monopole dependence with specific pole masses.}, decay
constants, as well as the light-cone wave functions of mesons
involved. This result is rather puzzling, and it might imply that
the naive factorization for $B$ decays to the $S$-wave charmonia
may still make sense but the overall normalization for the decay
rates are questionable.

In summary, we have studied the exclusive decays of $B$ meson into
pseudoscalar charmonium states $\eta_c$ and $\eta_c^\prime$ within
the QCD factorization approach and find that the nonfactorizable
corrections to naive factorization are infrared safe at
leading-twist order. The spectator interactions arising from the
kaon twist-3 effects are formally power-suppressed but chirally
and logarithmically enhanced. The theoretical decay rates are too
small to accommodate the experimental data. We already knew that
for $B\rightarrow J/\psi K$ decay, there are also logarithmic
divergences arising from spectator interactions due to kaon
twist-3 effects and the calculated rates are also smaller than
data by a factor of 8-10\cite{chay,cheng}. Moreover, in our
previous paper\cite{song}, we found that for $B \rightarrow
\chi_{c1} K$ decay, the factorization breaks down due to
logarithmic divergences arising from nonfactorizable spectator
interactions even at twist-2 order, and the decay rates are also
too small to accommodate the data, and that for $B\rightarrow
\chi_{c0} K$ decay, there are infrared divergences arising from
nonfactorizable vertex corrections as well as logarithmic
divergences due to spectator interactions even at leading-twist
order.

Considering the above problems encountered in describing
$B\rightarrow \eta_c K (\eta_c^\prime K)$ as well as $B\rightarrow
J/\psi K(\psi^\prime K)$ and $B\rightarrow \chi_{cJ}K(J=0,1)$
decays, we would like to restate our conclusion that in general
the QCD factorization method with its present version can not be
safely applied to exclusive decays of $B$ meson into charmonia,
and that new ingredients or mechanisms should be introduced to
describe exclusive decays of $B$ meson to charmonium states.

\section*{Acknowledgements}
We thank G. Bodwin, E. Braaten, X. Ji and C.S. Lam for useful
discussions and comments. This work was supported in part by the
National Natural Science Foundation of China, and the Education
Ministry of China.


\begin{thebibliography}{99}

\bibitem{BBNS1} M. Beneke, G. Buchalla, M. Neubert and
C.T. Sachrajda, Nucl. Phys. B  {\bf 591 } (2000)313.

\bibitem{BBNS2} M. Beneke, G. Buchalla, M. Neubert and
C.T. Sachrajda, Phys. Rev. Lett. {\bf 83 } (1999)1914.

\bibitem{BBNS3} M. Beneke, G. Buchalla, M. Neubert and
C.T. Sachrajda, Nucl. Phys. B  {\bf 606 } (2001)245.

\bibitem{chay} J. Chay and C. Kim, hep-ph/0009244.

\bibitem{cheng}
H.Y.~Cheng and K.C.~Yang, Phys.\ Rev.\ D {\bf 63} (2001)074011.

\bibitem{song} Z. Song and K.T. Chao, hep-ph/0206253.

\bibitem{cleo} K.W. Edwards et al. (CLEO Collaboration),
Phys. Rev. Lett. {\bf 86 } (2001)30.

\bibitem{babar}  B. Aubert et al. (BaBar Collaboration),
hep-ex/0203040.

\bibitem{belle1} F. Fang et al. (Belle Collaboration),
hep-ex/0208047.

\bibitem{belle2} S.-K. Choi et al. (Belle Collaboration),
Phys. Rev. Lett. {\bf 89 } (2002)102001, Erratum-ibid.{\bf 89 }
(2002)129901.

\bibitem{BBL}  G. Buchalla, A.J. Buras and M.E. Lautenbacher,
Rev. Mod. Phys. {\bf 68} (1996)1125.

\bibitem{ball} P. Ball, JHEP {\bf 9809} (1998)005.

\bibitem{constant} N. G. Deshpande, J. Trampetic, Phys. Lett. B {\bf 339} (1994)270.

\bibitem{pdg} D.E. Groom et al. (Particle Data Group), Eur. Phys. J. C {\bf 15} (2000)1.

\bibitem{chao} K.T. Chao, Y.F. Gu and S.F. Tuan,
Commun. Theor. Phys. {\bf 25} (1996)471.

\bibitem{gu} Y.F. Gu, Phys. Lett. B {\bf 538} (2002)6.

\bibitem{keum2}
M. Gourdin, Y.Y. Keum and X.Y. Pham, Phys.\ Rev.\ D {\bf 52}
(1995)1597.

\bibitem{keum1}
M. Gourdin, Y.Y. Keum and X.Y. Pham, Phys.\ Rev.\ D {\bf 51}
(1995)3510.


\end{thebibliography}
\end{document}